\documentclass[twocolumn,showpacs,preprintnumbers,amsmath,amssymb]{revtex4-1}

\usepackage{graphicx}
\usepackage{dcolumn}
\usepackage{bm}

\begin{document}

\title{Ultra-cold Fermi gases with resonant dipole-dipole interaction}
\author{T. Shi$^{1,3}$, S.-H. Zou$^{1}$, H. Hu$^{2}$, C.-P. Sun$^{1}$, and S. Yi$^{1}$}
\affiliation{$^1$Institute of Theoretical Physics, Chinese Academy of Sciences, P.O. Box
2735, Beijing 100190, China}

\affiliation{$^{2}$ARC Centre of Excellence for Quantum-Atom Optics, Centre for Atom Optics and Ultrafast Spectroscopy, Swinburne University of Technology, Melbourne 3122, Australia}

\affiliation{$^{3}$Max-Planck-Institut f\"{u}r Quantenoptik, Hans-Kopfermann-Str. 1, 85748 Garching, Germany} 

\date{\today }

\begin{abstract}
The superfluid phases in the resonant dipolar Fermi gases are investigated by the standard mean-field theory. In contrast to the crossover from Bose-Einstein condensation (BEC) to Bardeen-Cooper-Schrieffer (BCS) superfluid in the Fermi gases with the isotropic interactions, the resonant dipolar interaction leads to two completely different BEC phases of the tight-binding Fermi molecules on both sides of the resonance, which are characterized by two order parameters with the distinct internal symmetries. We point that near the resonance, the two competitive phases can coexist, and an emergent relative phase between the two order parameters spontaneously breaks the time-reversal symmetry, which could be observed in the momentum resolved rf-spectroscopy.
\end{abstract}

\maketitle

{\em Introduction}. --- The unprecedented experimental progresses~\cite{krb,lics,rydb,nak,dy} in creating quantum gases of fermions with large dipole moment has stimulated extensive investigations of dipolar Fermi gases. Owing to the long-range and anisotropic nature of the dipole-dipole interaction (DDI), new quantum phenomena emerge in the dipolar Fermi gases, e.g., the ferro-nematic phase \cite{ferro}, the novel static and dynamical properties in the normal phase \cite{np}, and the \textit{p}-wave dominated BCS superfluids induced by the partially attractive DDI \cite{you,bara,bara2,pu}. Of particular interest, the recent studies demonstrated that in the two-species dipolar Fermi gases the competition between the short-range contact interaction and DDI led to the coexistence of singlet- and triplet-paired superfluids~\cite{samo,shi,wu}.

The above theoretical work focus on the Fermi gases with DDI in the weak coupling regime, for which the scattering amplitude is replaced by the first Born approximation of the bare DDI potential~\cite{FBA}. Recent experimental progresses toward creating ultracold gases of polar molecules~\cite{krb,lics,nak} and Rydberg atoms~\cite{rydb} provide an opportunity for exploring the dipolar effects in the strong coupling regime. Previous studies on the low-energy scattering between two polarized dipoles indicate that, due to the shape resonances, the scattering lengths diverge at certain strengths of the DDI~\cite{mari,you,blume,zhai}. In particular, these shape resonances may take place simultaneously in multiple scattering channels as the DDI couples partial waves with different angular momenta~\cite{blume}. This {\em multichannel resonance} (MCR), described by a matrix of the scattering lengths, is in striking contrast to the Feshbach resonance (see, e.g., Ref.~\cite{chin} and references therein).

In this Letter, we investigate the novel superfluid phases in a two-species dipolar Fermi gas with resonant DDI. Based on the analysis to the low-energy scattering of two polarized dipoles, we propose an effective two-body interaction potential and a model to describe the MCR. It is found that a two-body bound state is formed on either side of the resonance. Consequently, for a many-body system, the Bose-Einstein condensates of the tight-binding molecules dominates on both sides of the resonance. Since these two molecular states possess distinct internal symmetries, the system experiences a phase transition across the MCR. Moreover, near the resonance, the two competitive phases coexist and a relative phase between the two order parameters emerges, which spontaneously breaks the time-reversal symmetry (TRS). Similar mixed order parameters were studied in the high-$T_{c}$ superconductors~\cite{mixed,mixed1,mixed2,mixed3,mixed4}. Finally, we study the quasi-particle spectral function of the system in order to explore the possibility of the experimental detection using the momentum-resolved (MR) rf-spectroscopy~\cite{rf,rf2}.

{\em Model}. --- We consider an ultracold gas of two-species dipolar fermions in a box of volume ${\cal V}$. For simplicity, we assume that $n_{\uparrow}=n_{\downarrow}=n$ with $n_{\alpha}$ being the number density of the spin-$\alpha$ particle. The total Hamiltonian of the system can be decomposed into $H=H_{0}+H_{1}$, where, in the momentum space, $H_{0}=\sum_{\mathbf{k},\alpha }\varepsilon _{\mathbf{k}}c_{\mathbf{k}\alpha }^{\dagger }c_{\mathbf{k}\alpha }$
with $c_{\mathbf{k}\alpha }$ being the annihilation operator of the spin-$\alpha$ fermion and $\varepsilon _{\mathbf{k}}=\mathbf{k}^{2}/(2M)-\mu$ with $M$ being the mass of the particle and $\mu$ the chemical potential. Furthermore, the interaction Hamiltonian takes the form
\begin{equation}
H_{1}=\frac{1}{2\mathcal{V}}\sum_{\alpha \beta }\sum_{\mathbf{kpq}
}U(\mathbf{k}-\mathbf{p})c_{\frac{\mathbf{q}}{2}\mathbf{+k}\alpha }^{\dagger
}c_{\frac{\mathbf{q}}{2}\mathbb{-}\mathbf{k}\beta }^{\dagger }c_{\frac{
\mathbf{q}}{2}\mathbb{-}\mathbf{p}\beta }c_{\frac{\mathbf{q}}{2}+\mathbf{p}
\alpha },  
\label{Hint}
\end{equation}
where $U(\mathbf{k})=4\pi d^{2}(3\cos^{2}\theta_{\mathbf{k}}-1)/3$ is the Fourier transform of the bare DDI between two polarized (along $z$-axis) dipoles with $d$ being the dipole moment and $\theta_{\mathbf k}$ the angle between $\mathbf k$ and the $z$-axis. We note that the short-range interaction can be straightforwardly included in the two-body interaction potential $U$~\cite{shi,FBA}.

{\em Effective interaction potential}. --- In order to obtain valid results in the strong coupling regime, the bare DDI has to be renormalized. To this end, we consider the scattering between two polarized dipoles. In terms of the $K$-matrix, ${\cal K}(k)$, the scattering amplitude can be generally expressed as
\begin{eqnarray}
f({\mathbf{k}}^{\prime },{\mathbf{k}})|_{k=k^{\prime }}&=&4\pi\sum_{lml^{\prime }m'}
i^{l'-l}k^{-1}\left(\frac{1}{\mathcal{K}^{-1}-i}\right)_{lm}^{l'm'}\nonumber\\
&&\times Y_{lm}(\hat{\mathbf{k}})Y_{l^{\prime }m'}^{\ast }(\hat{\mathbf{k}}^{\prime }),
\label{f}
\end{eqnarray}
where the subscripts and superscripts of the bracket denote the matrix elements. Specifically, for DDI, since $U\propto Y_{20}$, ${\cal K}_{lm}^{l'm'}$ is nonzero only when $|l-l'|\leq 2$. In addition, because the DDI conserves the projection of the angular momentum, ${\cal K}$ is diagonal with respect to the magnetic quantum number $m$. Of particular importance, in the zero energy limit, the matrix elements ${\cal K}_{lm}^{l'm}(k)/k$ are all nonvanishing~\cite{mari,you}, which results in finite scattering lengths $a_{ll'}^{(m)}=-\lim_{k\rightarrow0}{\cal K}_{lm}^{l'm}(k)/k$.

To proceed further, we construct a matrix ${\cal A}$ whose elements are defined by the scattering lengths as $\mathcal{A}_{ll^{\prime }}^{(m)}=i^{l^{\prime }-l}a_{ll^{\prime }}^{(m)}$. Assuming that the matrix $\mathcal{A}$ is diagonalized in the orthonormal basis $w_{jm}(\hat{\mathbf{k}})=\sum_{l}d_{jl}Y_{lm}(\hat{\mathbf{k}})$ with the corresponding eigenvalues $\lambda_{jm}$. In this new basis, the scattering amplitude Eq. (\ref{f}) can be reexpressed as
\begin{equation}
f({\mathbf{k}}^{\prime },{\mathbf{k}}
)|_{k=k^{\prime }\rightarrow 0}=4\pi\sum_{jm}f_{jm}w_{jm}(\hat{\mathbf{k}})w_{jm}^{\ast }(
\hat{\mathbf{k}}^{\prime }),
\end{equation}
where $f_{jm}=-1/(\lambda_{jm}^{-1}+ik)$. It is clear that the eigenvalue $\lambda_{jm}$ represents the effective scattering length in the scattering channel $w_{jm}(\hat{\mathbf k})$. We can now construct a separable effective interaction potential as
\begin{equation}
U'(\hat{\mathbf{k}},\hat{\mathbf{k}'})=4\pi \sum_{jm}g_{jm}w_{jm}(\hat{\mathbf{k}})w_{jm}^{\ast }(\hat{\mathbf{k}'}),
\label{pp}
\end{equation}
where the coupling constants $g_{jm}$ satisfy the renormalization condition
\begin{equation}
\frac{1}{g_{jm}}+\int \frac{d^{3}q}{(2\pi )^{3}}\frac{M}{q^{2}}=\frac{M}{4\pi \lambda_{jm}}.
\end{equation}
The correctness of the effective potential $U'$ can be verified as it reproduces the two-body scattering amplitude Eq. (\ref{f}), therefore, it is valid in both weak and strong coupling regimes. We point out that two other pseudo-potentials applicable to the strong DDI were also proposed previously~\cite{andrei,dwwang}.

{\em Two-body physics in the resonance regime}. --- With the effective potential $U'$, it can be shown that if $\lambda_{jm}$ is positive, the scattering channel $w_{jm}(\hat{\mathbf k})$ may support a two-body bound state with binding energy $E_{b,jm}=-1/(M\lambda _{jm}^{2})$  (see, e.g. Ref~\cite{atomcol}). Indeed, numerical calculations show that as one tunes the dipole moment $d$ of the colliding particles, the scattering lengths $a_{ll'}^{(m)}$ appear to have various resonances~\cite{mari,you,blume}. Those shape resonances indicate the DDI indeed supports bound states. Noting that, within the same $l$ and $l'$ manifold, $a_{ll'}^{(m)}$ with $m=0$ is always the largest term~\cite{you}, here and henceforth, we restrict our analysis to $m=0$ and drop the index $m$.

To be specific, we consider the collision between a spin-$\uparrow$ and a spin-$\downarrow$ particles in the spin-singlet channel. In general, the shape resonances induced by the DDI are well-separated~\cite{blume}, which allows us to focus on a particular resonance, say the resonance of $a_{00}$ at dipole moment $d_{r}$.  Interestingly, Kanjilal and Blume found that a resonance on $a_{02}$ also takes place at the same position $d_{r}$~\cite{blume}. In principle, because the DDI couples the partial waves $|l-l'|\leq 2$, resonances should occur on all $a_{ll'}$ with $l,l'={\rm even}$. Therefore, unlike the Feshbach resonance, the shape resonance induced by the DDI may occur simultaneously in multiple scattering channels. Utilizing the fact that the widths of the resonances decrease with increasing $l+l'$~\cite{blume}, we propose a minimal model for the MCR by assuming that it is described by the matrix
$$\mathcal{A}_{\mathrm{sd}}=\left(
\begin{array}{cc}
a_{00} & -a_{02} \\
-a_{02} & 0
\end{array}
\right),$$
where all other scattering lengths are assumed to be zero. We note that the exact behavior of ${\cal A}_{\rm sd}$ depends on the details of the short-range physics of the colliding particles. However, as will be shown, some general properties for the MCR can be obtained by analyzing this simplest model.

The matrix ${\cal A}_{\rm sd}$ can be easily diagonalized to yield the eigenvalues $\lambda _{1,2}=[a_{00}\pm {\rm sgn}(a_{02})\sqrt{a_{00}^{2}+4a_{02}^{2}}]/2$ and the corresponding eigenstates $w_{1,2}(\hat{\mathbf k})=[s_{1,2}Y_{00}(\hat{\mathbf k})+Y_{20}(\hat{\mathbf k})]/\sqrt{s_{1,2}^{2}+1}$, where $s_{1,2}=-(y\pm \sqrt{y^{2}+4})/2$ with $y=a_{00}/a_{02}$. Clearly, independent of the values of the scattering lengths, only one of the eigenvalues can be positive, i.e., there only exists a single bound state for any given set of ($a_{00},a_{02}$). More specifically, on the upper ($a_{00}^{-1},a_{02}^{-1}$) parameter plane, we have $\lambda_{1}>0$, which leads to the binding energy $E_{b}=-1/M\lambda _{1}^{2}$ and the angular distribution of the bound state wave function $\vert w_{1}(\hat{\mathbf k})\vert ^{2}$. On the other hand, $\lambda_{2}$ is positive on the lower ($a_{00}^{-1},a_{02}^{-1}$) plane. Consequently, the angular distribution of the bound state wave function becomes $\vert w_{2}(\hat{\mathbf k})\vert ^{2}$. Since $w_{2}(\hat{\mathbf k})$ is orthogonal to $w_{1}(\hat{\mathbf k})$, the bound state changes its microscopic symmetry when $a_{02}$ changes its sign. One can also carry out a similar analysis when $a_{00}$ changes its sign. Therefore, we expect that a quantum phase transition arises as the system crosses an MCR.

We remark that the above discussion can be easily generalized to the MCR between two spin polarized particles by replacing the even orbital angular momentum quantum numbers with odd ones. In this case, the simplest model for the MCR is characterized by the matrix $\mathcal{A}_{\mathrm{pf}}=\left(
\begin{array}{cc}
a_{11} & -a_{13} \\
-a_{13} & 0
\end{array}
\right)$. The properties of the two-body bound state can be obtained straightforwardly.

{\em Many-body physics across an MCR.} --- Now, we turn to investigate the many-particle properties of an ultra-cold gas of two-species fermionic dipoles within the regime of an MCR. At zero temperature, the gap and phase fluctuations can be ignored. The system can then be treated by using the standard BCS mean-field theory. 

\begin{figure}[tbp]
\includegraphics[width=3.4in]{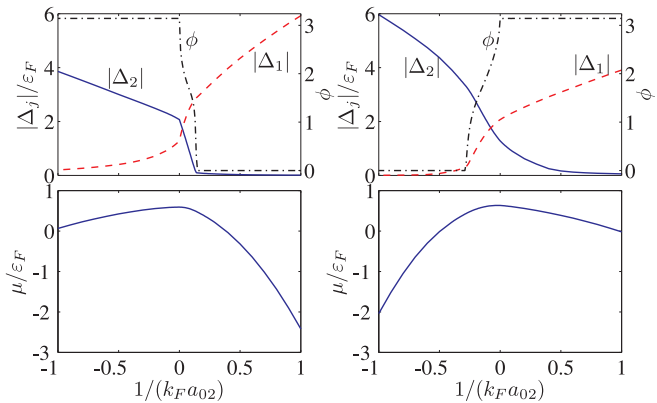}
\caption{(Color Online) Order parameters (upper row) and chemical potential (lower row) for a two-species dipolar Fermi gas in the resonance regime with $y=-1$ (left column) and $1$ (right column). Here, $\varepsilon_{F}=k_{F}^{2}/(2M)$ is the Fermi energy.}
\label{fig1}
\end{figure}

To start, we define the order parameters for the channels $w_{j}(\hat{\mathbf{k}})$ as $\Delta _{j}=4\pi g_{j}\sum_{\mathbf{k}}w_{j}^{\ast }(\hat{\mathbf{k}})\left\langle c_{-\mathbf{k}
\downarrow }c_{\mathbf{k}\uparrow }\right\rangle /\mathcal{V}$, which satisfy the coupled gap equations
\begin{equation}
-\frac{M\Delta _{j}}{16\pi ^{2}\lambda _{j}}=\sum_{j^{\prime }}\int \frac{
d^{3}p}{(2\pi )^{3}}w_{j}^{\ast }(\hat{\mathbf{p}})\left(\frac{1}{2E_{\mathbf{p}}}
-\frac{M}{p^{2}}\right)w_{j^{\prime }}(\hat{\mathbf{p}})\Delta _{j^{\prime }},
\label{GE}
\end{equation}
where $E_{\mathbf{p}}=\sqrt{\varepsilon_{\mathbf{k}}^{2}+\vert \Delta (\hat{\mathbf{k}})\vert ^{2}}$ is the quasi-particle excitation energy with $\Delta (\hat{\mathbf{k}})=\sum_{j}\Delta_{j}w_{j}(\hat{\mathbf{k}})$ being the order parameter. Furthermore, to completely determine the gaps $\Delta_{j}$, one needs the density equation, which, for the spin-balanced system, takes the form
\begin{equation}
\frac{k_{F}^{3}}{3\pi ^{2}}=\int \frac{d^{3}p}{(2\pi )^{3}}\left(1-\frac{\varepsilon _{\mathbf{p}}}{E_{\mathbf{p}}}\right),  \label{DE}
\end{equation}
where $k_{F}=(6\pi ^{2}n)^{1/3}$ is the Fermi momentum. Equations (\ref{GE}) and (\ref{DE}) form a closed set of equations for the order parameters $\Delta_{j}$ and the chemical potential $\mu$. Even though, the order parameters $\Delta_{j}$ are generally complex numbers, it can be shown that only their absolute values $|\Delta_{j}|$ and the relative phase $\phi={\rm arg}(\Delta_{1})-{\rm arg}(\Delta_{2})$ are relevant. 

For a given set of scattering lengths $a_{00}$ and $a_{02}$, the self-consistent Eqs. (\ref{GE}) and (\ref{DE}) can be solved numerically. In general, there exist multiple solutions corresponding to different local minima of the energy density, 
\begin{equation}
{\cal E}(\Delta_{1},\Delta_{2},\mu)=\int \frac{d^{3}k}{(2\pi )^{3}}\left[\varepsilon _{\mathbf{k}}-E_{\mathbf{k}
}+\frac{\vert \Delta (\hat{\mathbf{k}})\vert ^{2}}{2E_{\mathbf{k}}
}\right]+\frac{\mu k_{F}^{3}}{3\pi ^{2}}.\nonumber
\end{equation}
The true ground state can be identified by comparing $\cal E$ corresponding to the different solutions.

In principle, to obtain a full picture of the superfluidity in the resonant dipolar gases, one should numerically find the gaps for the scattering lengths covering the entire ($a_{00}^{-1},a_{02}^{-1}$) plane. Here, for simplicity, we fix the ratio $y=a_{00}/a_{02}$ of the scattering lengths and treat $a_{02}^{-1}$ as a free parameter. The assumption that $y$ is fixed implies that the widths of the resonances in $a_{00}$ and $a_{02}$ are the same, which seems rather unrealistic, as the resonance in $a_{02}$ is generally more narrow than that in $a_{00}$. However, the physics present here stands as long as the widths of the two resonances become comparable.

In Fig.~\ref{fig1}, we present the order parameters and the chemical potential of the system across an MCR for $y=\pm 1$. As can be seen, away from the resonance, the condensation of the molecular state [$\Delta_{1}$ ($\Delta_{2}$) for $a_{02}>0$ ($<0$)] always dominates. From the angular distributions of the order parameter $\Delta(\hat{\mathbf{k}})$, the Cooper pairs shows completely different internal symmetries on the different sides of the resonance. In this sense, a phase transition takes place as the system crosses the resonance. This result is in contrast to the Feshbach resonance, for which only a smooth crossover is experienced~\cite{jin,kett}. Near the resonance, $a_{02}^{-1}\sim 0$, the amplitudes of $\Delta_{1}$ and $\Delta_{2}$ become comparable such that the two order parameters coexist in the system. Of particular interesting, a nonzero relative phase also develops in this region. As a result, $\Delta(\hat{\mathbf k})$ becomes complex, which indicates that the TRS is spontaneously broken. In fact, such mixed order parameters with nonzero relative phases have been extensively studied in the high-$T_{c}$ superconductors~\cite{mixed}. 

Further evidence about the phase transition is provided by the chemical potential which  becomes negative away from the resonance. This indicates that strongly coupled BECs form on the both sides of the resonance. In addition, $\mu$ reaches its maximal value when the system approaches the resonanance, which explicitly shows the competition between these two phases. As a comparison, we point out that for the BCS-BEC crossover induced by the Feshbach resonance, the chemical potential monotonically decreases from the Fermi energy to a negative value as the system passes through the resonant regime from the BCS side to the BEC side.

It is also worthwhile to note that the sign of $a_{00}$ determines the detailed behavior of the order parameter for the BCS state. For negative (positive) $a_{00}$, the amplitude of the BCS state quickly (smoothly) decays to near zero when the system moves away from the resonance.


Replacing the matrix ${\cal A}_{\rm sd}$ by ${\cal A}_{\rm pf}$, one can generalize the above analysis to a single component dipolar Fermi gas in the MCR regime. Indeed, it is found that other than the angular distribution of the order parameter, the behaviors of the gaps and the chemical potential are very similar to those of a two-component gas.

\begin{figure}[tbp]
\includegraphics[width=3.2in]{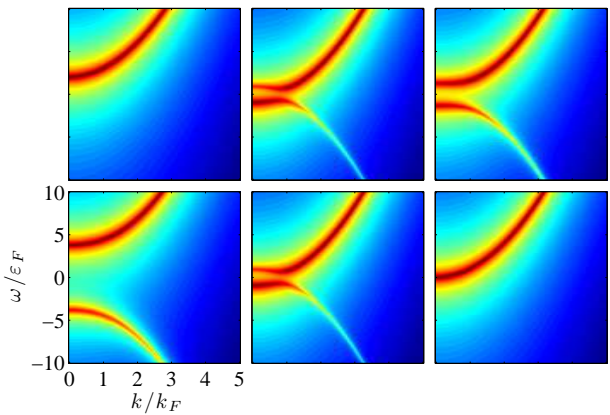}
\caption{(Color Online) Quasi-particle spectral function, $\ln A({\mathbf k},\omega)$, at $\cos\theta_{\mathbf k}=\pm0.39$ (upper row) and $\pm0.9$ (lower row) for a superfluid in the multiple-channel resonant regime. From the left to the right columns, $1/(k_{F}a_{02})=-1$, $0$, and $1$, respectively. Other parameters are $y=1$ and $
\gamma/\varepsilon_{F}=0.2$.}
\label{rfspec}
\end{figure}

{\em Experimental detection.} --- Now, we explore the possibility for the experimental detection of the fermionic superfluid with MCR. A potential technique for detecting the phase with mixed order parameters and the associated phase transition is the MR rf-spectroscopy~\cite{rf}. In such experiment, an rf pulse drives the transition from one of the two spin states to an unoccupied third spin state, and the photoemission spectroscopy measures the quasi-particle spectral function
\begin{equation}
A(\mathbf{k},\omega )=\frac{\gamma }{\pi }\sum_{s=\pm }\frac{u_{\mathbf{k},s}^{2}}{(\omega -sE_{\mathbf{k}})^{2}+\gamma ^{2}},
\end{equation}
where $u_{\mathbf{k},\pm}^{2}=(1\pm \varepsilon _{\mathbf{k}}/E_{\mathbf{k}})/2$ and $\gamma $ is the energy resolution. 

\begin{figure}[tbp]
\includegraphics[width=2.8in]{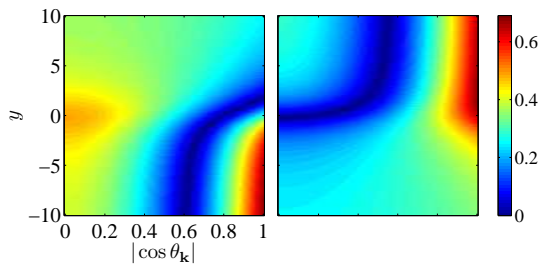}
\caption{(Color Online) $y$ dependence of the angular distributions for $|w_{1}(\theta_{\mathbf k})|$ (left panel) and $|w_{2}(\theta_{\mathbf k})|$ (right panel).}
\label{angle}
\end{figure}

As an example, we present $\ln A({\mathbf k},\omega)$ in Fig.~\ref{rfspec} for the ratio of the scattering lengths $y=1$. At the left end of the resonance ($1/k_{F}a_{02}\sim -1$), the system is dominated by the order parameter $\Delta_{2}w_{2}({\mathbf k})$. It can be shown that the gap is closed at $\cos\theta_{\mathbf k}^{*}\simeq\pm 0.39$ and reaches its maximum value at $\cos\theta_{\mathbf k}^{**}\simeq\pm0.9$. This properties is clearly demonstrated in the left column of Fig.~\ref{rfspec}. On the other hand, for $1/k_{F}a_{02}\sim 1$, the order parameter roughly becomes $\Delta_{1}w_{1}({\mathbf k})$, for which the gap is closed at $\theta_{\mathbf k}^{**}$ and opens up at $\theta_{\mathbf k}^{*}$. The behavior of the spectral function can then be used to identify the phase transition. Moreover, as shown in the middle column of Fig.~\ref{rfspec}, the gap presents for both $\theta_{\mathbf k}^{*}$ and $\theta_{\mathbf k}^{**}$ in the resonance regime ($1/k_{F}a_{02}\sim 0$). In fact, in this regime, it can be shown that the gap is nonzero for arbitrary $\theta_{\mathbf k}$.

For an arbitrary $y$, it is still possible to distinguish the two phases away from the resonance by using the spectral function. To demonstrate this, we plot the $y$ dependence of the angular distributions of the order parameters $w_{j}({\mathbf k})$ in Fig.~\ref{angle}. It can be shown that for $y<y_{1}\simeq1.79$ ($y>y_{2}\simeq-0.22$), $w_{1}$ ($w_{2}$) always has a zero at $|\cos\theta_{\mathbf k}|>0.58$ ($<0.58$). Therefore, for $y_{2}<y<y_{1}$, the spectral function possesses similar features as those displayed in Fig.~\ref{rfspec}. While for $y<y_{2}$ and $y>y_{1}$, only at one end of the resonance one can observe that the gap is closed. The angle $\theta_{\mathbf k}$ at which the gap vanishes can be used to identify the phase.

Finally, we note that the similar analysis for the spectral function of the MCR with odd angular momenta can be carried out straightforwardly, for which the gap always vanishes at $\cos\theta_{\mathbf k}=0$.

{\em Conclusions}. --- We have studied the superfluid phases of a two-species dipolar fermionic gas across an MCR. Based on the analysis to the low energy scattering of two polarized dipoles, we derive a separable effective potential for the DDI which is valid in the strong coupling regime. Subsequently, we proposed a minimal model which describes the MCR. We then studied the two-body and the many-body physics across an MCR. It was found that condensates of the molecular states with distinct internal symmetries form on both sides of the resonance, indicating that a phase transition arises when the system passes through the resonance. Near the resonance, the two competitive order parameters coexist. In addition, a relative phase between those two orders emerges, which spontaneously breaks the TRS. We also studied the quasi-particle spectral function of the system in oder to explore the possibility of the experimental detection using the MR rf-spectroscopy. Finally, we point out that due to the spontaneous TRS breaking, the resonant dipolar Fermi gas provide an opportunity for studying the Majorana edge modes in the spin-polarized dipolar Fermi gases~\cite{Ma}.

This work was supported by the NSFC (Grants No. 11025421, No. 10974209, and No. 10935010) and National 973 program under the Grants No. 2012CB922104. TS was partially supported by the EU project AQUTE.

\end{document}